\documentclass[superscriptaddress,showpacs,aps,twocolumn,prb,floatfix]{revtex4-1}
\usepackage{amssymb}
\usepackage{amsmath}
\usepackage[dvips]{graphicx}
\usepackage{bm}

\begin{document}
\title{Enhancing of nonlinear thermoelectric response of a correlated quantum dot in the Kondo 
regime by asymmetrically coupling to the leads}

\author{D. P\'erez Daroca}
\affiliation{Gerencia de Investigaci\'on y Aplicaciones, Comisi\'on Nacional de
Energ\'ia At\'omica, (1650) San Mart\'{\i}n, Buenos Aires, Argentina}
\affiliation{Consejo Nacional de Investigaciones Cient\'{\i}ficas y T\'ecnicas,
(1025) CABA, Argentina}

\author{P. Roura-Bas}
\affiliation{Centro At\'{o}mico Bariloche, Comisi\'{o}n Nacional
de Energ\'{\i}a At\'{o}mica, 8400 Bariloche, Argentina}
\affiliation{Consejo Nacional de Investigaciones Cient\'{\i}ficas y T\'ecnicas,
(1025) CABA, Argentina}

\author{A. A. Aligia}
\affiliation{Centro At\'{o}mico Bariloche and Instituto Balseiro, Comisi\'{o}n Nacional
de Energ\'{\i}a At\'{o}mica, 8400 Bariloche, Argentina}
\affiliation{Consejo Nacional de Investigaciones Cient\'{\i}ficas y T\'ecnicas,
(1025) CABA, Argentina}

\begin{abstract}
We study the low temperature properties of the differential response of the current to a temperature gradient
at finite voltage in a single level quantum dot including electron-electron interaction, non-symmetric
couplings to the leads and non-linear effects. 
The calculated response is significantly enhanced in setups with large asymmetries between the tunnel
couplings. In the investigated range of voltages and temperatures with corresponding energies up to several
times the Kondo energy scale, the maximum response is enhanced nearly an order of magnitude with respect to
symmetric coupling to the leads.
\end{abstract}

\pacs{73.23.-b, 71.10.Hf, 75.20.Hr}

\maketitle

\section{Introduction}

\label{intro}

During the last decade the study of nanodevices that convert heat into work has received great 
attention due to possible applications \cite{casati}. These quantum thermoelectric machines are 
usually considered to be composed by a system of a small number of degrees of freedom coupled to  
a set of macroscopic reservoirs. Among them, semiconducting \cite{gold,cro,gold2,wiel,grobis,kreti,ama} 
and molecular \cite{liang,kuba,yu,leuen,parks,roch,scott,parks2,serge,vincent} quantum dots (QDs)
play a fundamental role.

A permanent interest of enhancing efficiency and power of such devices exists. In this sense, the  
charge and heat currents through QDs have been focused of intense research both 
experimentally \cite{cui,guo,kim,rincon} and
theoretically \cite{cui,mah-so,krawiec,ala,costi,flensberg,linke,andergassen,crepieux,lomba2,corna,ang,lomba1,dorda,erdman,klo,sierra,jiang,roura-arrachea}. 

Mahan and Sofo have shown that a delta shape of the transmission function maximizes the efficiency 
of thermoelectric devices \cite{mah-so}. Since then, other mechanisms for increasing the Seebeck 
coefficient (thermopower), $S$, have been proposed.  
Among them, a time-dependent gate voltage \cite{crepieux}, orbital degeneracy \cite{lomba2,corna}, negative 
values of the Coulomb repulsion \cite{andergassen}, quantum Hall bar with fractional filling factors 
\cite{roura-arrachea}, and nonlinear transport effects \cite{flensberg,jiang,dorda} have been 
investigated.

Recently, Dorda \textit{et al.} studied  the differential thermoelectric response of a correlated impurity 
in the nonequilibrium Kondo regime at finite voltage.\cite{dorda} In addition to providing a fundamental 
understanding of the system, the authors point out the
potential of quantum dots as possible nanoscale temperature
sensors. The study was limited to symmetric coupling of the dot to both leads, and the authors state 
that further studies including asymmetric couplings to the
leads are required to fully assess the potential of quantum
dot devices for nanoscale sensing applications.

In this work we consider a QD asymmetrically coupled to the leads 
and we show that for large asymmetry, which is expected for molecular QD \cite{haug,rak}, the differential
response of the current to a temperature gradient at finite voltage is increased nearly by an order of magnitude 
with respect to the symmetric case.

\begin{figure}[tbp]
\includegraphics[clip,width=7cm]{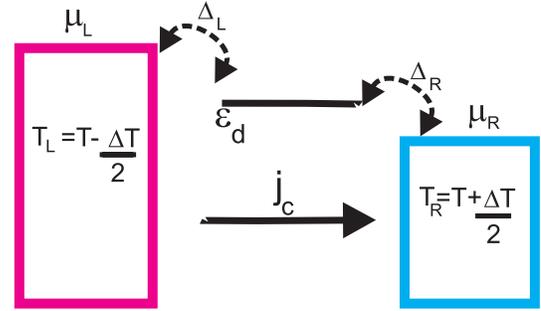}
\caption{(Color online) Scheme of the model considered, with the electronic level of the QD and their 
hybridization to the leads. 
}
\label{esquema}
\end{figure}

Specifically, we investigate nonlinear (NL) transport effects on the differential 
thermopower.  
When a bias voltage is applied to the system a constant electric current in the steady state 
is established. Assuming this regime, we analyzed its response to an infinitesimal gradient of 
temperature between the leads,
as sketched in Fig. \ref{esquema}. 
We restrict our study to the case in which transport is due to electrons. 
Vibrational effects were studied in Refs. \onlinecite{klo,rincon,flensberg}, among others.
The model is suitable for molecular QDs (for which asymmetric couplings are the most usual situation)  
and semiconducting QDs in which the tunnel couplings can be tuned 
at will \cite{ama}.
We find that the asymmetry of the tunneling couplings, $\alpha=\Delta_L /\Delta_R$,   
plays a nontrivial role and large values of $\alpha$ boost the differential thermopower in the NL regime.
As it will be clarified below, the range of parameters for this enhancement are within or near the 
Kondo regime.

\section{Model and expression for the current}

\label{model}

We employ the single impurity Anderson Hamiltonian to model the  molecular 
or semiconducting QD.
It is composed by a single dot level of energy $E_{d}$ connected to two metallic reservoirs. 
The Hamiltonian is 
\begin{eqnarray}
H &=& E_{d}n_{d}+Un_{d\uparrow
}n_{d\downarrow }+\sum_{\nu k\sigma }\epsilon _{k}^{\nu }c_{\nu k\sigma
}^{\dagger }c_{\nu k\sigma }  \notag \\
&&+\sum_{\nu k\sigma }(V_{k}^{\nu }d_{\sigma }^{\dagger }c_{\nu k\sigma }+
\mathrm{H.c}.)  \label{ham}
\end{eqnarray}
where $n_{d}=\sum_{\sigma }n_{d\sigma }$, $n_{d\sigma }=d_{\sigma }^{\dagger
}d_{\sigma }$, $d_{\sigma }^{\dagger }$ creates an electron with spin 
$\sigma $ at the active level of the QD, 
$c_{\nu k\sigma }^{\dagger }$ creates a
conduction electron at the left ($\nu =L$) or right ($\nu =R$) lead, and 
$V_{k}^{\nu }$ describe the hopping elements between the leads and the
QD.

When the temperatures $T_\nu$ and/or chemical potentials $\mu_\nu$ of the reservoirs are 
different, heat and electric (charge) currents flow from one lead to the other. The sign of such 
currents depend on the temperature and chemical potentials differences and they are constant in the 
steady state. As a reference, we assume $T_\nu=T+\gamma_{\nu}\Delta T/2$  and 
$\mu_\nu=-e\gamma_{\nu}\Delta V/2$ 
with the sign $\gamma_{\nu}=-(+)$ for $L(R)$ being $\Delta T>0$ the temperature difference and $\Delta V$ 
the bias voltage.

The charge current through the QD is given by \cite{meir} 
\begin{eqnarray}\label{currents}
J_{C}&=& \frac{2e\pi}{h}A(\alpha)\Delta \int d\omega 
 \rho (\omega )\left( f_{L}(\omega)-f_{R}(\omega),\right)
\end{eqnarray}
where $\Delta=\Delta_L + \Delta_R$ is the total coupling of the QD to the leads with
$\Delta _{\nu }=\pi \sum_{k}|V_{k}^{\nu }|^{2}\delta (\omega
-\epsilon _{k}^{\nu })$ (assumed independent of energy) and $A(\alpha)=4\alpha/(\alpha+1)^2$ represent the asymmetry factor.
In addition, $f_{\nu}(\omega)=1/(\exp(\frac{\omega-\mu_{\nu}}{T_{\nu}} + 1))$ is the Fermi distribution
associated to the lead $\nu$, and the spectral function of the QD per spin, is given by $\rho (\omega )$.

In the following, we assume $U\rightarrow \infty $, 
which is a realistic limit for most molecular QDs.
In fact, renormalizing the Kondo energy scale $T_K$ the physics discussed here is the same for finite $U$ as long 
as the Kondo peak (the one near the Fermi energy and of width $2T_K \ll \Delta$)  in the spectral density of states 
is well separated from 
the charge-transfer peaks at $E_d$ and $E_d+U$ and voltages and temperatures are such that the corresponding 
energy scales $eV$ and $T$ are much smaller than the charge-transfer excitations
($eV, T \ll |E_d|, E_d+U$ with the Fermi energy set as $\epsilon_F=0$). Since the total width
of the charge-transfer peaks in the Kondo regime 
$-E_{d}, E_d + U \gg \Delta $ are $\sim 4 \Delta$,\cite{capac,anchos} and $E_d$ can be tuned in 
QDs so that $E_d \sim -U/2$, our results are valid for systems such that $U \ge 8 \Delta$. 
 
\section{Noncrossing approximation}

\label{nca}

To calculate the spectral function of the QD, we use the non-crossing approximation (NCA) in its 
non-equilibrium extension \cite{win,roura_1}. 
The NCA is equivalent to a sum of an infinite series of diagrams in
perturbations in $V_{k}^{\nu} $.\cite{bickers,kroha} 
In the Kondo
regime, it is known to reproduce correctly the relevant energy scale $T_{K}$
and its dependence on the different parameters.
The out of equilibrium NCA approach is one of the standard techniques for calculating the  
current in mixed valence systems within different regimes of the model, and specially within 
the Kondo regime (KR) where the dot occupancy is near 1. 
It has proved to be a
very valuable tool for calculating the differential conductance
through a different systems such as two-level QD's and C$_{60}$ molecules
displaying a quantum phase transition, or a nanoscale Si transistor 
\cite{serge,roura_2,tosi_1,tetta} among others. It also reproduces correctly the scaling
of the conductance for small bias voltage $V$ and temperature $T$,\cite{roura_3}
and is able to reproduce finite-energy features in systems, where the numerical 
renormalization group has difficulties, like the presence of a step in the conduction band.\cite{ds}

Alternatives to NCA for non-equilibrium
problems have some limitations. For example renormalized perturbation theory is limited to small $\omega $, 
$V$ and $T$ \cite{ang,hbo,ogu,ct} and the method of the equation of motion,\cite{rome1,rapha,rome2,noise} 
does nor reproduce correctly the functional dependence of $T_{K}$ on $E_{d}$.\cite{rapha,rome2}

The main limitations of the NCA for the Anderson model we consider take place for
moderate positive $E_{d}$ and finite moderate $U$. For positive $E_d$, the impurity self energy has an
unphysical positive imaginary part and as a consequence $\rho (\omega )$ presents a spurious
peak at the Fermi energy. Similar spurious peaks exist for finite magnetic field.\cite{win}  
For finite $U$ the NCA ceases to reproduce correctly the dependence of $T_K$ with parameters 
and vertex corrections
should be included.\cite{pruschke89,haule01,tosi11} 
Since in this work, the parameters correspond to the Kondo regime $-E_{d}, E_d + U \gg \Delta $
and we take $U\rightarrow \infty $, we avoid these limitations. 
A minor problem is that the intensity of the Kondo peak is overestimated by about 15 \% 
compared to the value expected from the Friedel sum rule.\cite{su42}

More details on the formalism and tricks that we use to solve the selff-consistent integral equations can be found in Refs. 
\onlinecite{roura_1,benz}.

\section{Results}

\label{res}

Without loss of generality, we choose our unity of energy to be $\Delta = 1$.
We present results for $E_d=-4$, which corresponds to the KR.
We choose a bandwidth of $2D$ with $D=10$.  We define the Kondo temperature ($T_K$) 
as the temperature for which the equilibrium conductance is half of 
the corresponding one for the unitary limit, 
$G(T=T_K)=G_0/2$ with $G_0=2A(\alpha) e^2/h$. 
For our  parameters, this leads to $T_K=0.0086$. 
Due to the universality of the model in the KR the results presented here in units of $T_K$ are quite 
general in this regime and do not depend on the particular values of $\Delta$, $E_d$ or $U$.
For simplicity, we also take the Boltzmann constant and absolute value of the electronic charge
$k_B=e=1$ so that The Seebeck coefficient $S$ and $V/T$ become dimensionless.

The Seebeck coefficient, $S$, and the electrical conductance, $G$, within the linear response regime,
$\Delta T=\Delta V \rightarrow 0 $, are commonly defined in terms of equilibrium properties \cite{costi},
$S=-I_1(T)/[eTI_0(T)]$, $G=e^2 I_0(T)$ being 
\begin{equation}\label{I_n}
I_n(T)=\frac{2\pi}{h}A(\alpha)\Delta \int d\omega ~\omega^{n} ~ \rho (\omega )(-f'(\omega)).
\end{equation}
Here, the asymmetry factor $\alpha$ does not modify the equilibrium spectral density due to the fact that both,
the charge-transfer peak near $E_d$ and the Kondo one at the Fermi energy $\epsilon_F=0$
depend on the total coupling $\Delta=\Delta_L+\Delta_R$ and not on the ratio $\alpha=\Delta_L /\Delta_R$. 
While the charge-transfer peak has a width
of $4\Delta$ in the KR \cite{capac,anchos}, the width of the Kondo peak is of the order of 
$T_K\sim D\exp(\pi E_d/ 2\Delta)$ 
and its intensity is fixed by the Friedel sum rule $\rho(0)\sim 1/\pi\Delta$. 

\begin{figure}[tbp]
\includegraphics[clip,width=8cm]{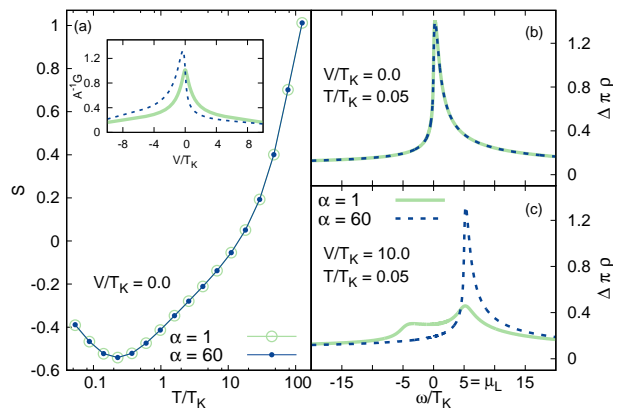}
\caption{(Color online) (a) Thermopower as a function of temperature for two values of the asymmetry ratio $\alpha$
at equilibrium ($\Delta V=\Delta T =0$). 
The inset shows the differential conductance as a function of 
bias voltage for two values of  $\alpha$ (1 and 60). 
(b) Spectral density as a function of energy at 
equilibrium for the same $\alpha$ as in (a). 
(c)  Spectral density as a function of energy for $V/T_K=10$ for the same $\alpha$ as in (a).
In all cases $T/T_K=0.05$.}
\label{fig2}
\end{figure}

In Fig. \ref{fig2}(b) we show the spectral density
for two extremely different values of $\alpha$. As explained above, they are identical. 
Thus, the equilibrium conductance is strongly suppressed for high values of the asymmetry  $\alpha$ 
[due to the factor  $A(\alpha))$ in Eq.  (\ref{currents})] 
while the Seebeck coefficient remains unchanged due to the cancellation of $A(\alpha)$ 
in  $I_1(T)/I_0(T)$ as it can be seen 
in the left panel of Fig.  \ref{fig2}.

However, when finite values of the bias voltage are considered, non trivial effects arise, which depend on $\alpha$. 
The Kondo resonance in the spectral density splits in two peaks, located at $\mu_{\nu}$ \cite{win,st-jpcm}.
The weights of both peaks are affected by $\alpha$. While they are approximately equal in the symmetric case, $\alpha=1$,
the one at $\mu_R$ ($\mu_L$) decreases (increases) for increasing $\alpha$. For  $\alpha \gg 1$, 
the QD is nearly in equilibrium with the left lead and the spectral density shows the full Kondo resonance shifted to
the chemical potential $\mu_L$, see Fig. \ref{fig2}(c).  
On the other hand, the differential conductance, $G(\Delta V)=dJ_C / d (\Delta V)$, is an even 
function of the bias voltage for $\alpha=1$ due to left-right reflection symmetry but in the case of $\alpha\gg1$ 
it mimics the spectral density, $G(\Delta V)\sim \frac{e^2}{h} \pi \Delta A(\alpha)\rho(-e\Delta V/2)$, 
see inset of Fig. \ref{fig2}(a).
Similar conclusions were drown in Refs. \onlinecite{capac, krawiec}.

In what follows we focus on the analysis of nonlinear effects on the differential Seebeck coefficient and 
its dependence on $\alpha$.
In analogy to bulk systems, when a temperature difference $\Delta T$ is applied between both sides of a QD, 
keeping $J_C=0$,
a voltage difference $\Delta V$ proportional to $\Delta T$ appears, 
$S = - \frac{\Delta V}{\Delta T}\big\vert_{J_C = 0}$.
Non linear effects following the line $J_C = 0$ have been already 
addressed \cite{jiang,lomba2,krawiec,lomba1,flensberg}.
However, Dorda \textit{et al.} have recently analyzed the non linear regime across the line $J_C \ne 0$,
\begin{equation}\label{seebeck}
 \mathcal{S} = - \frac{d(\Delta V)}{d(\Delta T)}\big\vert_{J_C = const.} = \frac{\partial J_C}{\partial(\Delta T)} \Big/
                                                                 \frac{\partial J_C}{\partial(\Delta V)}
\end{equation}
which is expected to be a measurable quantity that contributes to the better knowledge of the two decoherence 
processes generated by the bias voltage and temperature and can be important for applications in nanoscale 
temperature sensing \cite{dorda}. 

We compute numerically both $\partial J_C/\partial(\Delta V)$  from Eq. (\ref{currents}) and
$\partial J_C/\partial(\Delta T)$ using
\begin{eqnarray}\label{partial-respect-T}
 \frac{\partial J_C}{\partial(\Delta T)}&=& \frac{2e\pi}{h}A(\alpha)\Delta \int d\omega 
 \Big[ \frac{1}{2T}\rho(\omega)\sum_{\nu}\big( \omega-\mu_{\nu} \big)\frac{df_{\nu}}{d\omega}(\omega)\nonumber \\
  &+&\big( f_L(\omega)-f_R(\omega) \big) \frac{\partial \rho(\omega)}{\partial(\Delta T)} \Big].
\end{eqnarray}

We will show that the differential thermopower at finite bias voltage as described by
Eq.  (\ref{seebeck}) is largely enhanced for large asymmetry between the tunnel couplings. 

\begin{figure}[tbp]
\includegraphics[clip,width=8.0cm]{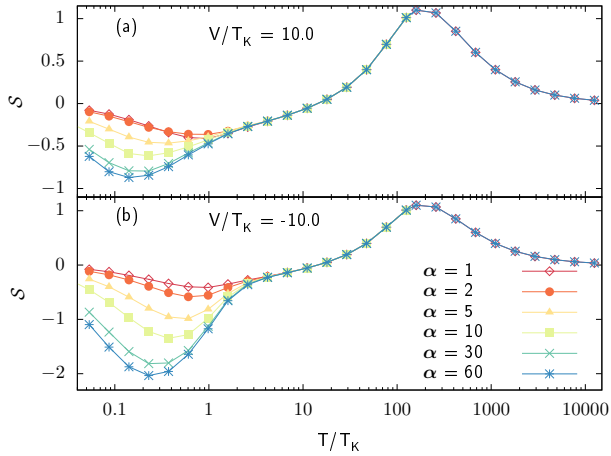}
\caption{(Color online) Differential thermopower as a function of temperature for several 
values of the asymmetry $\alpha$, (a) with $e\Delta V/T_K=10$ and (b) with $e\Delta V/T_K=-10$.}
\label{fig3}
\end{figure}

In Fig.  \ref{fig3} we show the results for the differential Seebeck coefficient at finite voltage
as a function of temperature and for several values of the asymmetry factor $\alpha$. 
For $T \gg T_K$, $\mathcal{S}$ does not depend on $\alpha$ independently of the sign 
of the current. In fact, at high $T$ there is no Kondo peak in the spectral density and the electronic 
transport is hole-like mediated by the charge transfer peak located at $\omega\sim E_d <0$, that has a width of 
the order of $4 \Delta$ \cite{capac,anchos}, and is nearly not affected for the bias voltage considered in the figure, 
of the order of a few $T_K$. For finite $U$, the other charge transfer peak at $\omega\sim E_d +U >0$, neglected in 
our approach can modify the results. However, we are interested in the regime of temperatures of 
the order of a few times $T_K$ or less. For these temperatures, only the spectral density at low energies is relevant due to 
the small window of the Fermi functions. One can observe a significant increment of the magnitude of $\mathcal{S}$ 
for larger values of $\alpha$
as compared to the symmetric case. In particular for $V=-10 T_K$ and temperatures of the order of $T_K$ or below it,
$|\mathcal{S}|$ is enhanced by a factor near 5 as the asymmetry $\alpha$ increases from 1 (symmetric case) to 60.

\section{Interpretation and further results}

\label{inter}

\begin{figure}[tbp]
\includegraphics[clip,width=8.0cm]{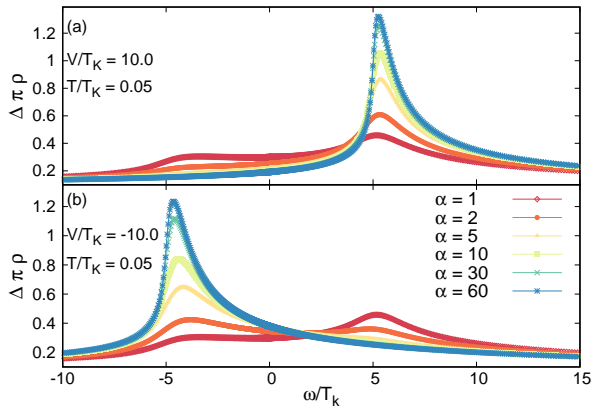}
\caption{(Color online) Kondo resonance in the QD spectral density under a finite bias voltage 
for several values of the asymmetry ratio $\alpha$. 
(a) Positive voltage, $\mu_L/T_K=5$, $T_K=0.008$. (b) Negative voltage, $\mu_L/T_K=-5$, $T_K=0.01$.}
\label{fig4}
\end{figure}

In order to understand the behavior of the thermoelectric response as $\alpha$ is varied, 
we analyze qualitatively the low-temperature features of the charge current in 
Eq.  (\ref{currents}).
As we have showed in Fig. \ref{fig2}, for large enough asymmetry the dot tends to be in equilibrium 
with the left lead. The evolution of the spectral density with $\alpha$ is explicitly shown in 
Fig.  \ref{fig4} for both signs of the applied bias voltage.

If the dot is in equilibrium with the left lead, then $\int d\omega \rho (\omega ) f_{L}(\omega)$
does not change with the applied voltage.
Using this assumption and   
standard Sommerfeld expansion \cite{rincon} 
for the current in Eq.  (\ref{currents}), the low temperature behavior of the differential Seebeck coefficient
reads as follows, 
\begin{eqnarray}\label{S-approx}
 &\mathcal{S}(T)&\sim  \big[\rho(\mu_R)\big]^{-1}\Big[-\frac{\pi^2 T}{6e}\sum_{\nu}\rho'(\mu_{\nu})\nonumber\\ 
     &&  + \int_{\mu_{R}}^{\mu_{L}} d\omega h(\omega)+ \frac{\pi^2 T^2}{6e}\big( h'(\mu_{L})-h'(\mu_{R}) \big)\Big],
\end{eqnarray}
being $h=\partial \rho/\partial \Delta T$ and $A'\equiv dA/d\omega$.
Fig.  \ref{fig4} demonstrates that $\sum_{\nu}\rho'(\mu_{\nu})$ in the first term of the right-hand side 
of Eq.  (\ref{S-approx})
increases as the coefficient $\alpha$ does. 
On the other hand, the value of $\rho(\mu_R)$ continuously decreases when $\alpha$ is increased.
Interestingly, both tendencies contribute to boost the coefficient $\mathcal{S}$. 

\begin{figure}[tbp]
\includegraphics[clip,width=8.0cm]{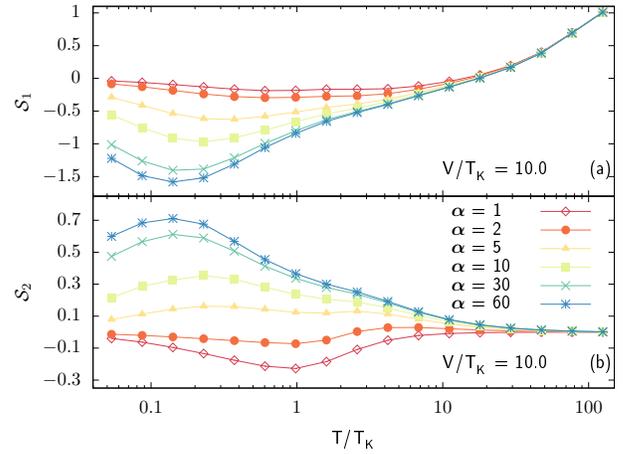}
\caption{(Color online) Contributions to $\mathcal{S}$ of the first and second members of  
Eq.  (\ref{partial-respect-T}), $\mathcal{S}_1$ and $\mathcal{S}_2$ respectively, 
as a function of temperature and for positive bias voltage.}
\label{fig5}
\end{figure}

While the temperature dependence of $\mathcal{S}$ is qualitatively the same for positive as well as negative 
bias voltage, as it is shown in panel (a) ($\Delta V>0$) and (b) ($\Delta V<0$) of Fig.  \ref{fig3}, 
the different intensity between them is related to the relative sign of the different contributions in 
Eq.  (\ref{S-approx}). 
We have verified that for $\alpha>5$ the magnitude $h(\omega)$ and $h'(\omega)$ are positive within the 
relevant energy 
range, and therefore, the second term of 
Eq.  (\ref{S-approx}) is also positive. 

In Fig.  \ref{fig5} we separately show for a positive value of voltage, the first and second contributions to $\mathcal{S}$ from 
Eq.  (\ref{partial-respect-T}), namely $\mathcal{S}_1$ and $\mathcal{S}_2$ respectively. 
While the total magnitude of $\mathcal{S}$ is still larger for the asymmetric case, it is clear that both 
contributions partially compensate each other.
However, for negative voltage (not shown) both terms have the same negative sign and boost even further the magnitude of 
$\mathcal{S}$. The chain in 
sign is clearly due to the third (last) term in Eq.  (\ref{S-approx}).
Therefore, larger intensity of the differential response is obtained for $\mu_L < 0$, that is when the 
current flows from 
the less coupled (right) lead to the other one.
This observation can be used to determine experimentally, in a very simple way, to what side ($L$ or $R$) 
the molecule or QD is closer or more coupled.

\begin{figure}[tbp]
\includegraphics[clip,width=8.0cm]{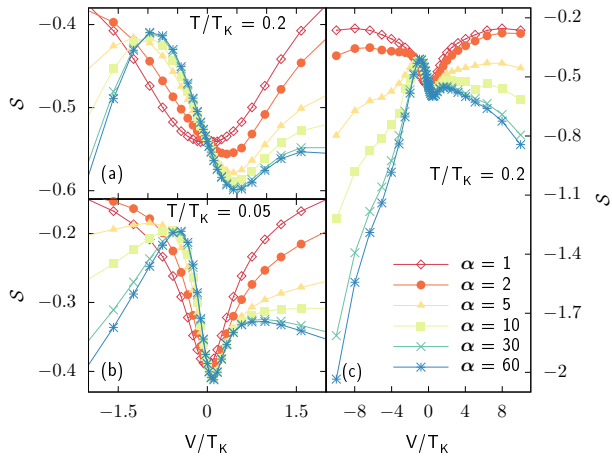} 
\caption{(Color online) Differential thermopower as a function of the bias voltage in units of $T_K$ 
for several values of the asymmetry ratio $\alpha$ and at low temperature.}
\label{fig6}
\end{figure}

In Fig.  \ref{fig6} we show the behavior of the thermopower at two different temperatures and for several values 
of the asymmetry factor $\alpha$ as a function of the applied bias voltage.
As a first observation, for the bias voltage considered here, $ \vert e\Delta V\vert \sim T_K $, the sign
of $\mathcal{S}$ is negative in agreement with Fig.  \ref{fig3} for $T<T_K$, 
due to the electronic character of the transport. 

In the case of symmetric couplings, $\alpha=1$, $\mathcal{S}$ as defined in Eq.  (\ref{seebeck}) is characterized by an even 
function of the applied bias voltage\cite{s-even}. The minimum at $ e\Delta V = 0$ is due to finite temperature effects. 
From Eq.  (\ref{partial-respect-T}) 
$\mathcal{S}$ is expected to vanish in the zero temperature limit. This tendency can be observed by a comparison
of the minimum value of $\mathcal{S}$ for the two selected temperatures in Fig.  \ref{fig6}, $T=0.2 T_K$ and 
$T=0.05 T_K$.
On the other hand, for large values of $ e\Delta V $, the argument of Eq.  (\ref{partial-respect-T}) tends to 
be an odd function within the relevant range of energies. Note that the later can be written as  
$[\rho(\omega+e\Delta V/2)+\rho(\omega-e\Delta V/2)]~\omega f'(\omega)$ in case of the first contribution 
(See Fig.  \ref{fig2}(c)). Regarding the second one, we note that $h_{\nu}(\omega)$ has structure at both $\mu_{\nu}$ 
and therefore the product $(f_{L}(\omega)-f_{R}(\omega))h(\omega)$ approaches an odd function of the energy.

On the other hand, the case of $\alpha\gg1$ is quite different.
While the behavior of $\mathcal{S}$ for $\vert e\Delta V \vert / T_K \ll 1$ is still governed by temperature effects,
its dependence on $e\Delta V$ has no parity. The situation of finite values 
of the bias voltage, particularly at values of a few times $T_K$, 
is more involved.  
As soon as  $ \vert e\Delta V\vert $ reaches $T_K$, the absolute value of $\mathcal{S}$ is 
largely increased. 
Once again, the mechanism behind this behavior is the displacement of $\rho(\omega)$ 
towards $\omega\sim\mu_L$. 
The same mechanism that explained the behavior of $\mathcal{S}$ in Fig.  \ref{fig3} applies in the case of 
the voltage dependence. 
Large values of $\mathcal{S}$ are obtained for large enough asymmetry. 
In particular, the increment is even larger when negative
voltage is considered as explained before. Note that for negative $V$ an increase in $|\mathcal{S}|$ for more than 
an order of magnitude is obtained as $\alpha$ increases from 1 to 60.

\section{Conclusions}

\label{conc}

In summary, we have investigated theoretically the differential response of the electric current 
at finite bias voltage, when 
an infinitesimal gradient of temperature is applied to a system of a molecular or semiconducting quantum dot 
coupled asymmetrically to two conducting  leads. 
We have concentrated on the electronic contribution to the differential Seebeck coefficient $\mathcal{S}$, 
leaving aside phonon contributions.

We show that $\mathcal{S}$ is strongly enhanced 
for  a large asymmetry between the tunnel couplings. 
In particular, we find an enhancement of $\mathcal{S}$ by an order of magnitude at temperatures 
of the order of a fraction of the Kondo temperature $T_K$ and bias voltage $eV$  of the order of several 
times  $k_B T_K$. 
This becomes relevant for the standard cryogenic 
experimental conditions, for which the Kondo effect emerges.
In Section \ref{inter}, we have provided an explanation of the non-trivial mechanism 
behind our findings and we believe that it could be useful for experimental purposes. 
Our findings can also be important for nanoscale temperature sensing.

Although the calculation of the figure of merit ZT is beyond the scope of this work, 
our results suggest that molecular quantum dots, for which large tunneling asymmetries are
expected, or semiconducting quantum dots in which a large asymmetry can be tuned easily, would be the 
most efficient quantum machines operating as both, heat-engines or refrigerators.

\section*{Acknowledgments}

We are partially supported by CONICET, Argentina. 
A. A. A, was sponsored by PIP 112-201101-00832 of CONICET and PICT
2013-1045 of the ANPCyT.


\begin{thebibliography}{99}
\bibitem{s-even} In the case of $\alpha=1$, $\rho(\omega)$ is invariant under $\Delta V\rightarrow-\Delta V$ 
(and as consequence $\partial J_C/\partial(\Delta T)$), 
see Fig.  \ref{fig3}. The same applies to $\sum_{\nu}\big( \omega-\mu_{\nu} \big)\frac{df_{\nu}}{d\omega}(\omega)$.
Therefore, the current in Eq.  (\ref{currents}) is and odd function of the bias voltage while $G(\Delta V)$ is an even one.
Furthermore, both contributions of Eq. (\ref{partial-respect-T}) are invariants under $\Delta V\rightarrow-\Delta V$.
\end{thebibliography}
\end{document}